\documentclass[onecolumn, 11pt]{IEEEtran}
\usepackage{amsmath}
\usepackage{amssymb}
\usepackage{amsfonts}
\usepackage{graphicx}
\usepackage{epsfig}
\usepackage{subfigure}
\usepackage{psfrag}
\usepackage{cite}
\usepackage{latexsym}
\usepackage{url}
\usepackage{color}

\begin{document}
\title{A General Utility Optimization Framework for Energy Harvesting Based Wireless Communications}
\author{
Hang Li, Jie Xu, Rui Zhang, and Shuguang Cui\\
\thanks{H. Li and S. Cui are with the Department of Electrical and Computer Engineering, Texas A\&M University, College Station, Texas, 77843 USA (e-mail: david\_lihang@tamu.edu; cui@ece.tamu.edu). S. Cui is also a Distinguished Adjunct Professor at King Abdulaziz University in Saudi Arabia.}
\thanks{J. Xu and R. Zhang are with the Department of Electrical and Computer Engineering, National University of Singapore (e-mail: elexjie@nus.edu.sg; elezhang@nus.edu.sg). R. Zhang is also with the Institute for Infocomm Research, ASTAR, Singapore.}

}

\setlength{\textwidth}{7.1in} \setlength{\textheight}{9.7in}
\setlength{\topmargin}{-0.8in} \setlength{\oddsidemargin}{-0.30in}

\maketitle

\begin{abstract}
In recent years, wireless communication systems are expected to achieve more cost-efficient and sustainable operations by replacing conventional fixed power supplies such as batteries with energy harvesting (EH) devices, which could provide electric energy from renewable energy sources (e.g., solar and wind). Such EH power supplies, however, are random and instable in nature, and as a result impose new challenges on reliable communication design and have triggered substantial research interests in EH based wireless communications. Building upon existing works, in this article, we develop a general optimization framework to maximize the utility of EH wireless communication systems. Our framework encapsulates a variety of design problems, such as throughput maximization and outage probability minimization in single-user and multiuser setups, and provides useful guidelines to the practical design of general EH based communication systems with different assumptions over the knowledge of time-varying wireless channels and EH rates at the transmitters.
\end{abstract}

\setlength{\baselineskip}{1.3\baselineskip}
\newtheorem{definition}{\underline{Definition}}[section]
\newtheorem{fact}{Fact}
\newtheorem{assumption}{Assumption}
\newtheorem{theorem}{\underline{Theorem}}[section]
\newtheorem{lemma}{\underline{Lemma}}[section]
\newtheorem{corollary}{\underline{Corollary}}[section]
\newtheorem{proposition}{\underline{Proposition}}[section]
\newtheorem{example}{\underline{Example}}[section]
\newtheorem{remark}{\underline{Remark}}
\newtheorem{algorithm}{\underline{Algorithm}}[section]
\newcommand{\mv}[1]{\mbox{\boldmath{$ #1 $}}}

\section{Introduction}
Energy harvesting (EH) is expected to have abundant applications in future wireless communication networks to power transceivers by utilizing the environmental energy such as solar, thermal, wind, and kinetic energy. Since the renewable energy is in general clean and cheap, EH offers various benefits compared to conventional energy supplies such as fossil fuel based generators. For example, in cellular networks, solar panels and wind farms have been deployed to power base stations, thus lowering the expenses on energy bills as well as reducing the level of carbon dioxide emissions. Besides, in wireless sensor networks, EH has been considered as a good substitute of the traditional battery, prolonging the operation time to almost infinity, at least theoretically.

Despite the above advantages, the introduction of EH also imposes new challenges on the communication system design. Specifically, the random and intermittent characteristics of renewable energy impose a new type of EH constraints, i.e., the available energy at an EH communication node up to any time is bounded by its accumulatively harvested energy by then. This is in contrast to conventional communication systems with fixed energy sources, in which the available energy at any time is either unbounded or only limited by the remaining energy in the storage device (e.g., battery). In addition, wireless communication channels often fluctuate more substantially and dynamically than the practical EH rates (e.g., the channel changes on the order of milliseconds (ms) while the EH rate changes on the order of seconds or minutes), while channel fading is the main challenge faced in the design of reliable wireless communications. Due to both the new EH constraints and the multi-time-scale channel/EH-rate variations, it is a challenging problem to jointly optimize the communication scheduling and energy management in EH based wireless communications.

Given both the advantages and challenges mentioned above, EH based wireless communications have drawn significant research attentions in recent years. Various EH-oriented transmission policies have been proposed for different applications (e.g., see \cite{Gunduz2014} and the references therein). Building upon the prior works, we aim to develop a general utility optimization framework in this article to reveal further the fundamental limits and design principles of EH based wireless communications, and apply this framework to solve some selected problems in single-user and multiuser communication setups. Finally, we will discuss some future research directions that we deem worthy of further investigation.

\section{Single-User EH Based Communications: System Model and Optimization Framework}

\begin{figure}
\centering
 \epsfxsize=1\linewidth
    \includegraphics[width=9cm]{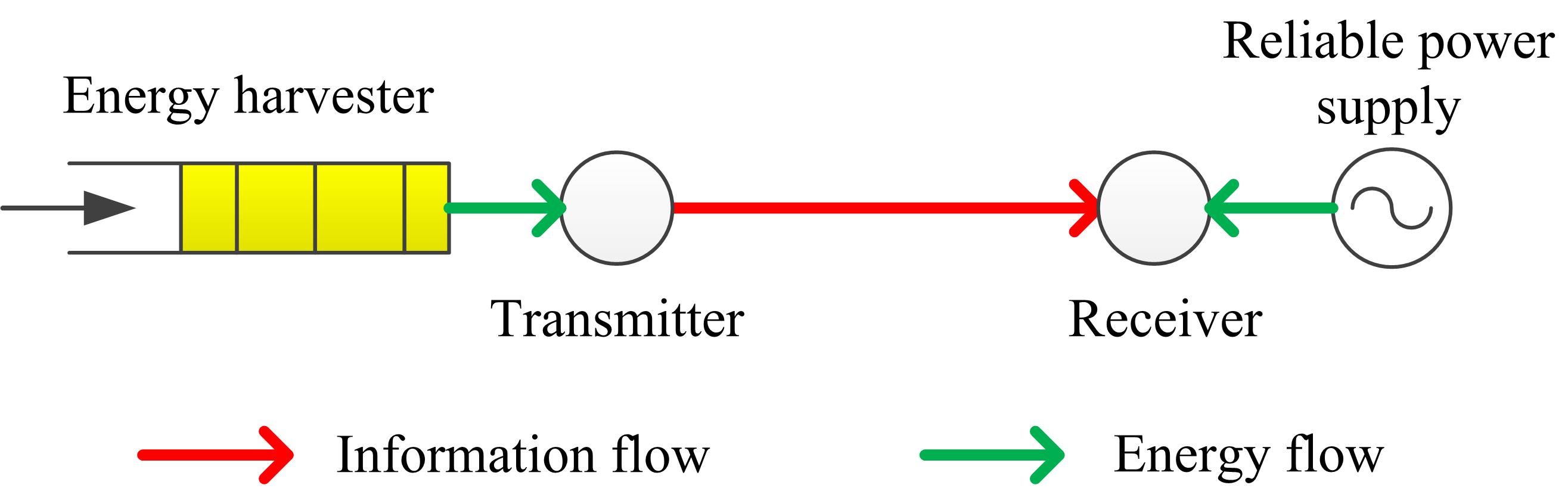}
\caption{A point-to-point communication link with an EH transmitter and a receiver with reliable power supply.} \label{p2pchannel}
\end{figure}

\begin{figure}
\centering
 \epsfxsize=1\linewidth
    \includegraphics[width=10cm]{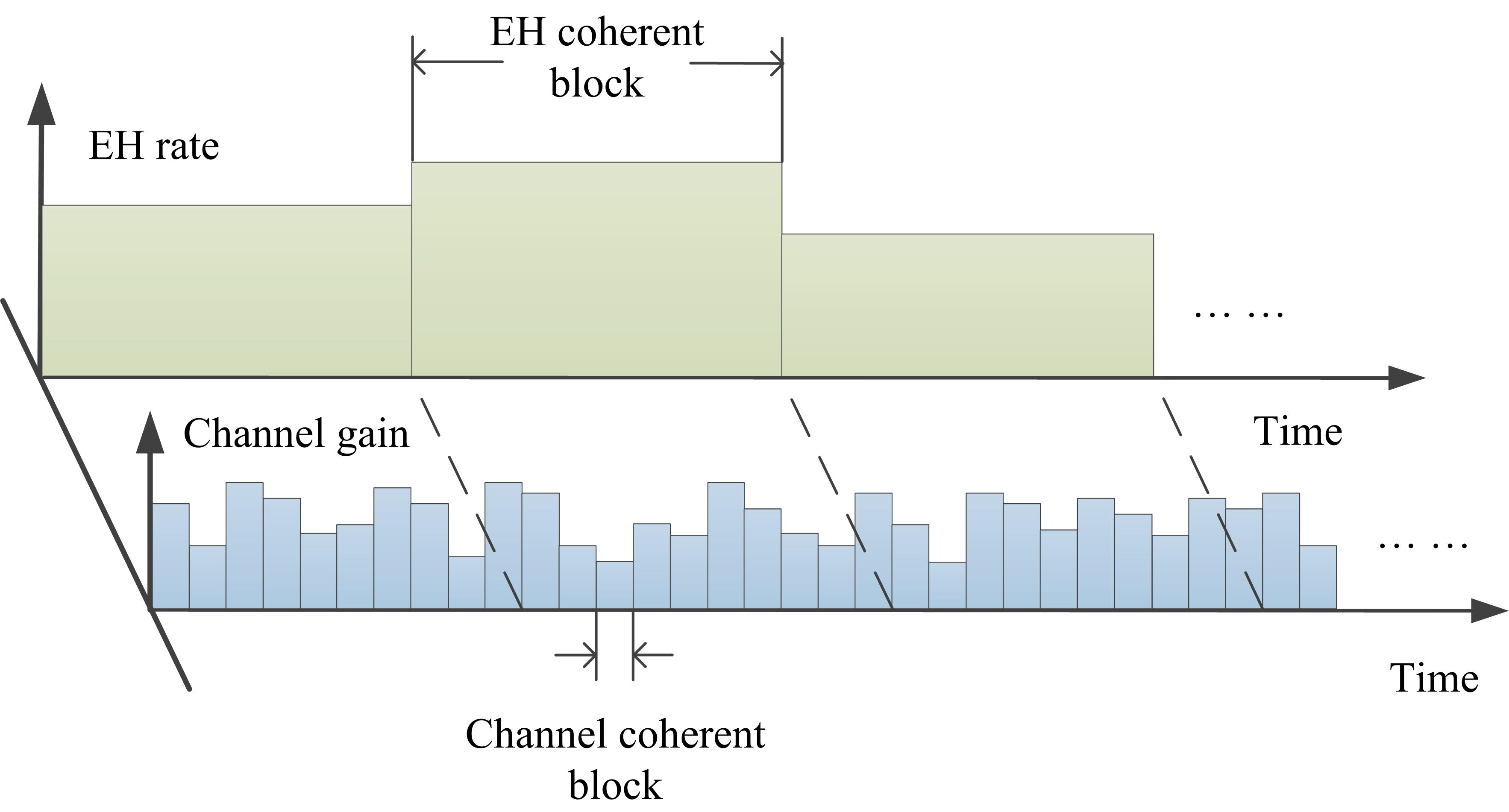}
\caption{Time-variation in EH process versus wireless channel.} \label{framework}
\end{figure}

First, we consider in this section a point-to-point wireless communication system as depicted in Fig. \ref{p2pchannel}, which consists of one transmitter powered by an energy harvester and one receiver with a reliable power supply. In practice, the coherence time of EH processes is often much larger than that of wireless channels as previously mentioned. Therefore, a block-based quasi-static EH model is practically valid, where the EH rate remains constant within each EH coherent block and may change from one block to another, and at the same time each EH block spans over many communication channel coherent blocks, as shown in Fig. \ref{framework}.

For the purpose of exposition, we consider wireless data transmissions over a finite horizon of $M\ge 1$ EH blocks. Each EH block is further divided into $N\ge 1$ communication blocks each of one unit time and a constant channel gain. Let $E(m) \ge 0$ denote the EH rate in the $m$-th EH block, and $h(n,m)\ge 0$ the channel power gain of the $(n,m)$-th communication block{\footnote{For notational convenience, we use $(n,m)$ to denote the $n$-th communication block of the $m$-th EH block.}} with $1\leq n\leq N$ and $1\leq m\leq M$. Furthermore, we use $P(n,m) \ge 0$ to denote the power consumption at the transmitter in the $(n,m)$-th communication block, which in general includes both the transmission power and the circuit power overhead. Assuming an ideal energy storage device (i.e., with infinite capacity and no energy leakage) employed at the transmitter, we have the {\it EH constraints} on the scheduled power consumptions $\{P(n,m)\}$; that is, the energy accumulatively consumed up to any communication block $(n,m)$, i.e., $\sum_{j=1}^{m-1}\left(\sum_{i=1}^N P(i,j)\right) + \sum_{i=1}^n P(i,m)$, should be no larger than the energy accumulatively harvested by then, i.e., $N\sum_{j=1}^{m-1} E(j) + n E(m)$ \cite{HoZhang2010,Ozel2011}.

To characterize the communication quality of service (QoS) measured at the receiver, we define a general utility function $U_{n,m}(P(n,m))$, which is dependent on the allocated power $P(n,m)$ at the transmitter for the $(n,m)$-th block. Note that in general the utility function $U_{n,m}(P(n,m))$ also depends on the channel power gain $h(n,m)$ of the $(n,m)$-th block. For example, the utility function can be defined more explicitly as throughout \cite{HoZhang2010,Ozel2011,XuZhang2014}, non-outage probability \cite{HuangZhangCui2014A,LuoZhangLim2013}, or other performance metrics such as end-to-end distortion in an EH based estimation system \cite{ZhaoZhang2013}, which could be either deterministic or statistical average based on the availabilities of the channel state information (CSI), i.e., $\{h(n,m)\}$, and the energy state information (ESI), i.e., $\{E(m)\}$, at the transmitter, namely, CSIT and ESIT, respectively.

\begin{figure}
\centering
 \epsfxsize=1\linewidth
    \includegraphics[width=10cm]{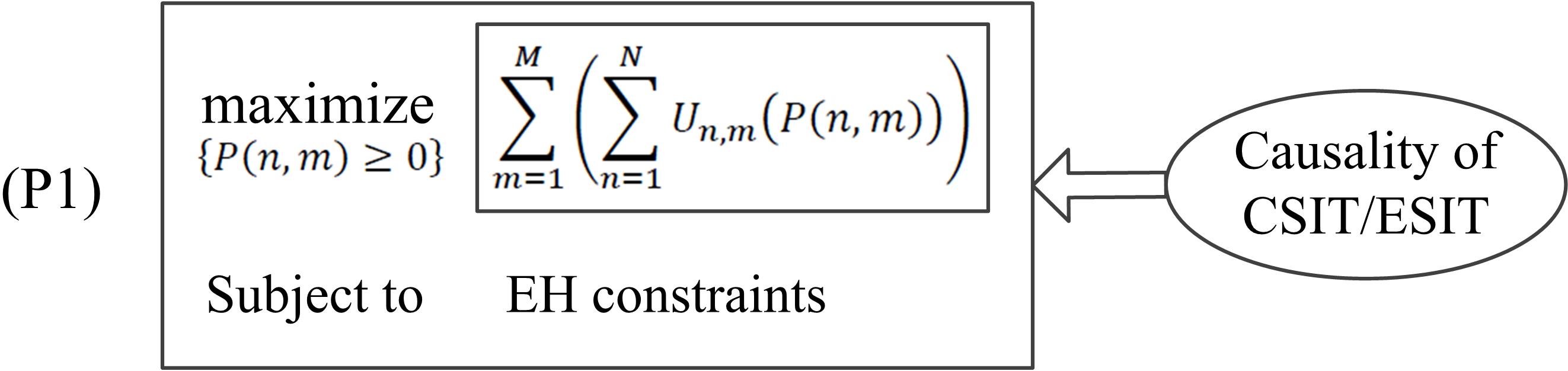}
\caption{The general utility maximization problem.} \label{Prob}
\end{figure}

Thus, the overall utility maximization problem over the $M$ EH blocks could be formulated as (P1) depicted in Fig. \ref{Prob}. Among all different assumptions about the CSIT and ESIT, there are four cases of our primary interest in this article as listed below:
\begin{itemize}
  \item Case 1: non-causal CSIT and ESIT. At the beginning of the transmission, the transmitter perfectly knows the past, current, and future CSI and ESI. This case approximates the practical scenario when the transmitter can accurately predict the future CSI (e.g., slowly-varying channels in low-mobility applications) and the future ESI (e.g., based on historical data in a periodically varying energy environment). The optimal solution in this case provides a performance upper bound for all other CSIT/ESIT availability cases.
  \item Case 2: causal CSIT and ESIT. At the beginning of each EH/communication block, the transmitter knows the past and current CSI/ESI, as well as the statistical information (e.g., distributions) of future CSI/ESI. In general, the solution of this case achieves the lowest utility among the first three cases considered herein.
 \item Case 3: causal CSIT and non-causal ESIT. This is a hybrid model based on Cases 1 and 2, in which all ESI is perfectly known at the beginning of the transmission, while only the past and current CSI is known.
 \item Case 4: no CSIT and non-causal/causal ESIT. During the transmission, the transmitter does not have any CSI, and only has the statistical information of the CSI. The non-causal or causal ESIT is defined as that in Case 1 or 2 above.
\end{itemize}
Note that in all the above cases, we assume that at each communication block, the receiver perfectly knows the CSI in that block.

\section{Optimal Solution for Single-User EH Based Communications}
In this section, we consider two example utility functions of the general utility maximization problem defined in problem (P1) in Fig. \ref{Prob}, namely, the throughput maximization and non-outage probability maximization (or equivalently, outage probability minimization). In the literature, these two utilities were commonly adopted to characterize the performance limits of communications over fading channels for delay-sensitive and delay-tolerant applications, respectively, assuming that the transmitter always has backlogged data to send. Based upon these two utilities, we reveal the structure of the optimal power allocation solution at the EH transmitter to problem (P1).

\subsection{Throughput Maximization}
For throughput maximization, the utility function $U_{n,m}(P(n,m))$ is generally modeled as the achievable rate at the $(n,m)$-th communication block, which is a positive non-decreasing and concave function of the transmit power $P(n,m)$ after ignoring the circuit power consumption at the transmitter (for the more general case with additional circuit power consumption considered, please refer to \cite{XuZhang2014,OzelUlukus2014}). For example, the Shannon capacity formula is widely used to model $U_{n,m}(P(n,m))$ \cite{HoZhang2010,Ozel2011,XuZhang2014,OzelUlukus2014}, i.e., $U_{n,m}(P(n,m))=\log_2\left(1+h(n,m)P(n,m)\right)$, in bits/sec/Hz. We present the optimal power allocation solution to problem (P1) as follows for different CSIT/ESIT cases introduced in the previous section.

First, consider the extreme case, i.e., Case 1, for obtaining the throughput upper bound. Given the non-causal CSIT and ESIT in this case, problem (P1) is a convex optimization problem since the objective function is concave and the constraints are all affine. By using the Karush-Kuhn-Tucker (KKT) conditions, it is shown in \cite{HoZhang2010,Ozel2011} that the optimal power solution exhibits a {\it staircase water-filling} structure, where the water level is a non-decreasing and staircase function over EH blocks. Particularly, if the additive white Gaussian noise (AWGN) channel model is assumed (i.e., $h(n,m)$'s are constant over all $n$'s and $m$'s), the transmitter should follow a non-decreasing staircase strategy for power allocation as depicted in Fig. \ref{throughput}.

As for Case 2, since both the CSI and ESI are only known causally at the transmitter, we are interested in maximizing the expected throughput over the randomness of the EH and channel variations, instead of its exact value as in Case 1. Note that due to the possible statistical correlation of EH rates and channel gains over time, the EH communication system may be considered as a dynamic system, for which some analytic models such as Markov decision process (MDP) and queueing system are applicable. The optimal power allocation could be derived based on the dynamic programming (DP) technique\footnote{Please refer to \cite{HoZhang2010} and \cite{Ozel2011} for more detailed discussions about the technique of DP.}  which balances the tradeoff between the exact throughput achieved in the current block versus the expected throughput over future blocks based on the knowledge of current CSIT and ESIT and their distributions. In general, DP may incur exponentially growing computation complexity as the number of EH/communication blocks increases. Therefore, suboptimal solutions with lower complexity have attracted a great deal of attention \cite{HoZhang2010,Ozel2011}, where certain throughput performance may be compromised to reduce the complexity.

For Cases 3 and 4, to our best knowledge, the throughput maximization problems have not been studied in the literature yet. Here, we provide brief discussions on these two cases to motivate future investigations. With causal CSIT and non-causal ESIT in Case 3, the transmitter is aimed to maximize the expected throughput over the randomness of channel realizations, subject to a set of deterministic EH constraints. In general, such a problem can be optimally solved by the DP technique similarly as in Case 2, based on which the transmitter needs to decide its power allocations with the updated CSI block by block. By exploiting DP and the deterministic EH constraints, it may be feasible to further obtain insightful and well-structured solutions, at least under some specific channel distributions, e.g., independent and identically distributed (i.i.d.) channel gains, which is an interesting problem worth pursuing.

In Case 4, the throughput optimization is generally a very challenging problem that remains open. The reason is that due to the lack of CSIT, the transmitter cannot adapt its transmit power and hence rate based on the instantaneous CSI, and therefore the receiver may fail to decode the information if the channel is too weak at that block. Thus, the total throughput may not be achievable in general. Despite this difficulty, one tractable case of this problem is when the number of communication blocks is sufficiently large within each EH block (say, $N\to \infty$). In this case, the ergodic capacity is achievable by letting the transmit power remain constant over each EH block and using a sufficiently long capacity-achieving code. With such a transmission scheme, the throughput utility achieved by each EH block can be expressed explicitly (although not in closed-form for general fading distributions); hence, problem (P1) can be formulated as an ergodic throughput maximization problem over $M$ EH blocks with non-causal or causal ESIT, which could be solved in the similar way to that for Cases 1 or 2.

\subsection{Outage Probability Minimization}

Outage occurs when there is a failure in decoding the data packet at the receiver, which is mainly due to the fact that the received signal undergoes a ``deep'' channel fading such that its power is not sufficient to combat against the receiver noise at the decoder. In EH based wireless communications, besides the channel fading, the uncertainty in the amount of harvested energy could be another source for transmission outage due to insufficient transmit power (or even insufficient circuit power; see \cite{LuoZhangLim2013}) at the EH transmitter.

To better address the outage issue, let $Q_{n,m}(P(n,m))$ denote the outage probability function of the $(n,m)$-th block, which depends on both the transmit power $P(n,m)$ and the channel power gain $h(n,m)$. For example, by assuming the Shannon capacity as the achievable rate, and that the required transmission rate at all $NM$ blocks is constant, the outage probability $Q_{n,m}(P(n,m))$ at the $(n,m)$-th block can be expressed as the probability that the achievable rate is less than the required transmission rate \cite{HuangZhangCui2014A}. Following this definition, the utility function $U_{n,m}(P(n,m))$ could be modeled by the non-outage probability for the $(n,m)$-th block, i.e., $U_{n,m}(P(n,m))=1-Q_{n,m}(P(n,m))$. It is thus straightforward to see that problem (P1) is equivalent to minimizing the average outage probability over the transmissions.

As the optimal solution structure for the previous throughput maximization crucially relies on the concavity of the rate function, the optimal power allocation for the outage probability minimization problem also critically depends on the properties of the outage probability function $Q_{n,m}(P(n,m))$. Here, $Q_{n,m}(P(n,m))$ is determined based on the availability of CSIT or the probability distribution of the channel gain $h(n,m)$, as will be specified for different cases in the next. For simplicity, we assume in this subsection that the channel gains $\{h(n,m)\}$ are i.i.d. over different communication blocks.

\begin{figure}
  \centering
  \subfigure[EH rates at the transmitter.]{\label{EHrates}\includegraphics[width=0.40\textwidth]{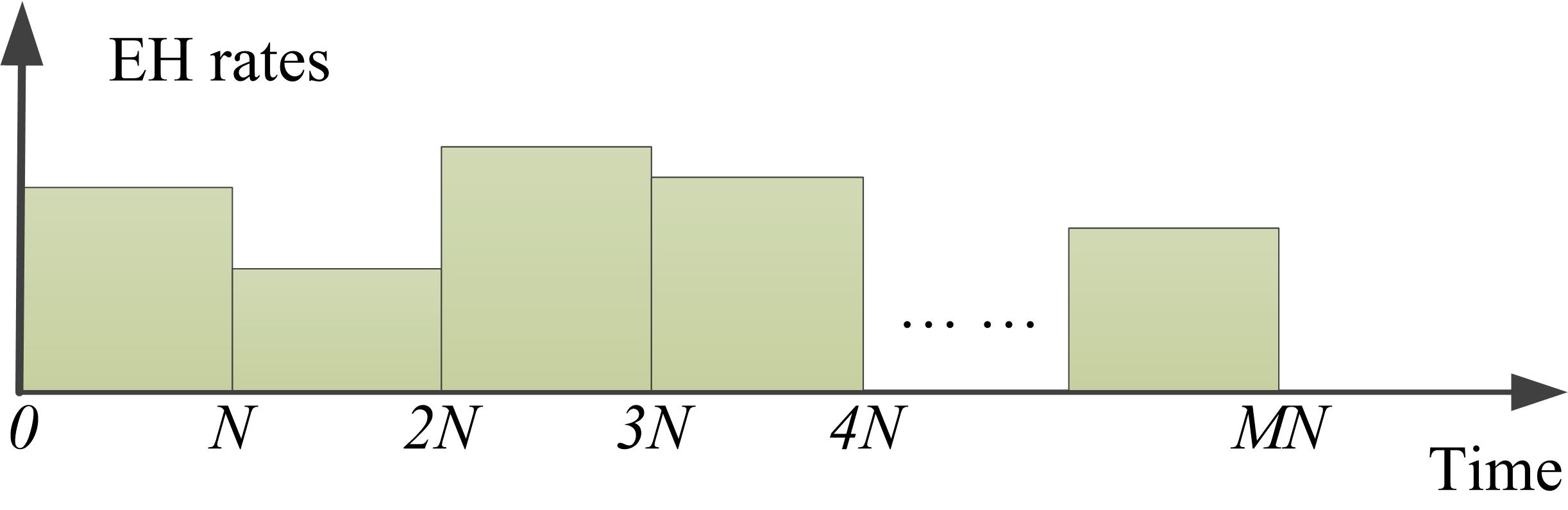}}\\
  \subfigure[The optimal power allocation for throughput maximization in Case 1 with an AWGN channel.]{\label{throughput}\includegraphics[width=0.40\textwidth]{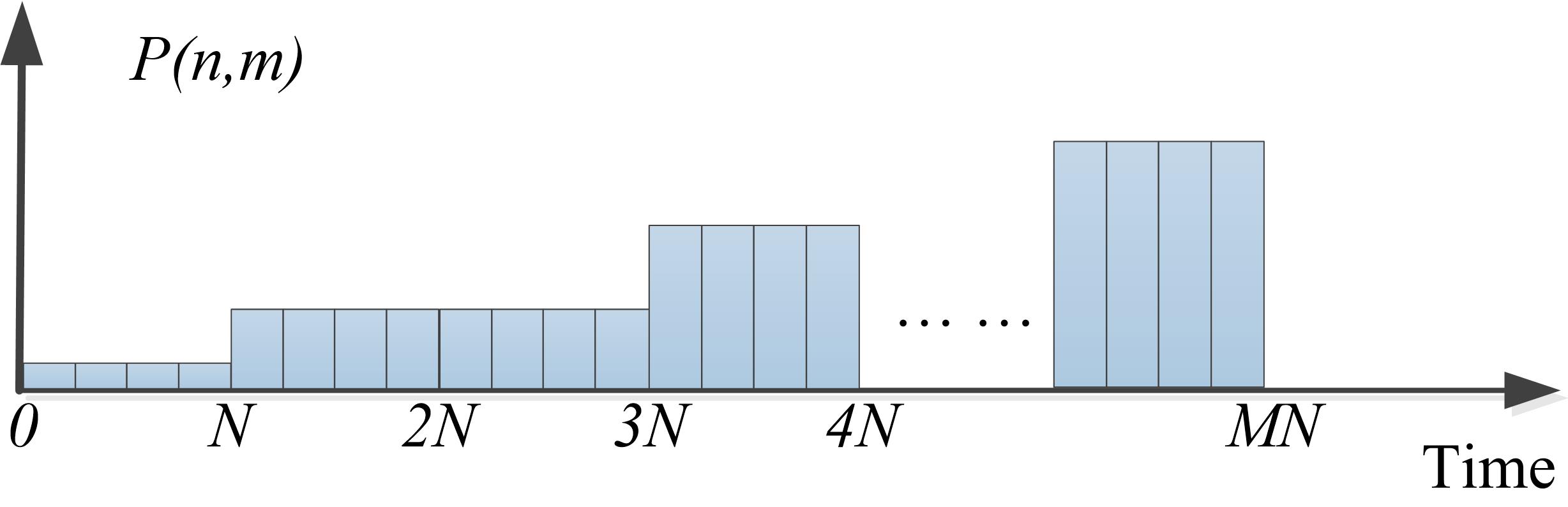}}\\
  \subfigure[The optimal power allocation for outage probability minimization in Case 3 with Weibull fading channel.]{\label{outage}\includegraphics[width=0.40\textwidth]{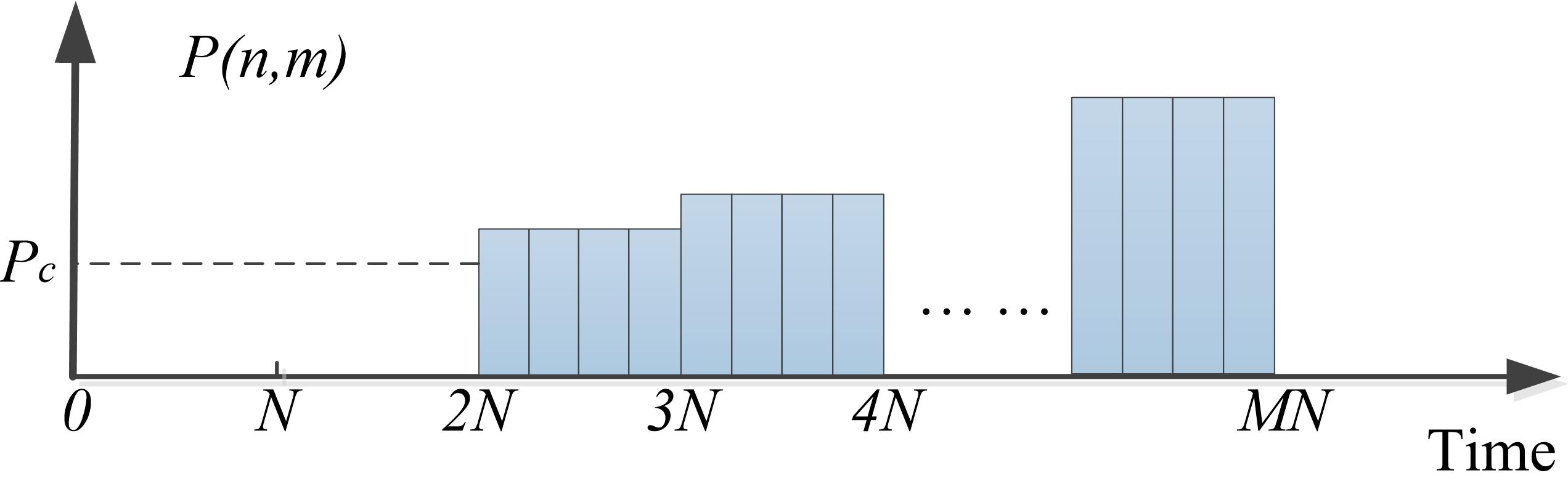}}
  \caption{Optimal transmit power allocations in a point-to-point channel subject to EH constraints.}
\end{figure}

First, we consider Case 4 without CSIT. The outage probability function $Q_{n,m}(P(n,m))$ is generally non-increasing over transmit power and satisfies one of the following two properties \cite{HuangZhangCui2014A}: 1) it is convex over transmit power on $[0,+\infty)$; or 2) there exists a critical point $P_c>0$ such that it is concave over $[0,P_c]$ and convex over $[P_c,+\infty)$. Such results hold for a large class of fading channels, including Weibull, Rician, Nakagami, and double Rayleigh fading, etc. Under the above properties of $Q_{n,m}(P(n,m))$, we study the optimal power allocation in Case 4 with non-causal ESIT. If the outage probability function satisfies Property 1), we can directly solve problem (P1) based on the similar techniques used for throughput maximization \cite{HoZhang2010,Ozel2011} as discussed in the previous subsection. However, if the outage probability function satisfies Property 2), the problem becomes non-convex, but is solvable by considering two regions: concave region $[0,P_c]$ and convex region $[P_c,+\infty)$. It is shown in \cite{HuangZhangCui2014A} that the optimal solution follows a ``save-then-transmit'' structure with nondecreasing power allocation as depicted in Fig. \ref{outage}. Specifically, at the beginning of the transmission, the transmitter keeps harvesting energy until the available power becomes larger than $P_c$, say, from time 0 to $2N$ in Fig. \ref{outage}. After that, the transmit power should be allocated non-decreasingly over time similar to the staircase water-filling.

For Case 4 with causal ESIT, problem (P1) becomes minimizing the expected outage probabilities over the randomness of EH rates, which can be solved by standard DP techniques \cite{HuangZhangCui2014A}. It is interesting to note that since EH rates vary per EH block spanning over $N$ communication blocks, the DP based optimal solution should obtain the optimal power allocation for all the $N$ constituting communication blocks at the beginning of the current EH block, by balancing the tradeoff between minimizing the outage probabilities in the current EH block versus those in the future EH blocks.

For the other three cases (i.e., Cases 1, 2 and 3), to our best knowledge, how to find the optimal power allocation for the outage probability minimization has not yet been investigated. In the following, we provide some intuitions that may be helpful to solve this problem under different cases.

Considering Case 1 with non-causal CSIT and ESIT, the outage probability function $Q_{n,m}(P(n,m))$ becomes an indicator function of the transmit power $P(n,m)$ for any given $(n,m)$-th block, i.e., $Q_{n,m}(P(n,m))=1$ if the achievable rate with power $P(n,m)$ is strictly less than the required transmission rate; otherwise, $Q_{n,m}(P(n,m))=0$. With this simplification, the average outage probability minimization is equivalent to minimizing the sum-value of the indicator functions over $1\le n\le N$ and $1\le m\le M$. Although such a problem is non-convex in general, it has the following structures. First, it is evident that the allocated transmit power at any communication block should be either zero for an outage event or the minimum required power for a non-outage transmission (with the resulting achievable rate equal to the required transmission rate), since the energy would be wasted otherwise. Another fact is that the objective value of each communication block is either zero or one. With these two observations, we conjecture that the optimal power allocation in this case could be found by first ordering the channel gains from the best to the worst, and then allocating the non-zero transmit power iteratively based on the order of channel gains subject to the EH constraints.

Finally, for Cases 2 and 3 with causal CSIT, the outage probability minimization corresponds to minimizing the expected outage probability over either the randomness of EH rates (Case 2) or deterministic EH constraints (Case 3). In general, the optimal power allocations for these two cases can be derived via standard DP techniques, and in each communication block, the transmitter should decide its current power allocation based on the binary objective value for outage. Interestingly, since the current CSIT is always known, the transmit power at each block $(n,m)$ should be either zero for an outage event or the minimum one for a non-outage transmission, similar to Case 1 above. This decision is made by balancing the tradeoff between avoiding the outage in the current block versus minimizing the future outage probabilities.

\section{Multiuser EH Based Communications}
So far, we have studied the optimal transmission policy for utility maximization in a single-user EH communication system. However, in practical wireless systems, multiple users with independent or shared energy sources may communicate over the same spectrum. In such systems, besides fading channels and time-varying EH rates of different users, their mutual interference among each other is a new challenge to be tackled in the multiuser utility optimization. In this section, we present promising communication and/or energy cooperation approaches among EH users to maximize the system-level utility. First, we consider the classic three-node relay channel with EH source and relay, and then we discuss other multiuser setups with EH transmitters.

\subsection{Energy and Communication Cooperation in Relay Channel}
\begin{figure}
  \centering
  \subfigure[Joint power allocation without energy sharing]{\label{withoutEHsharing}\includegraphics[width=0.50\textwidth]{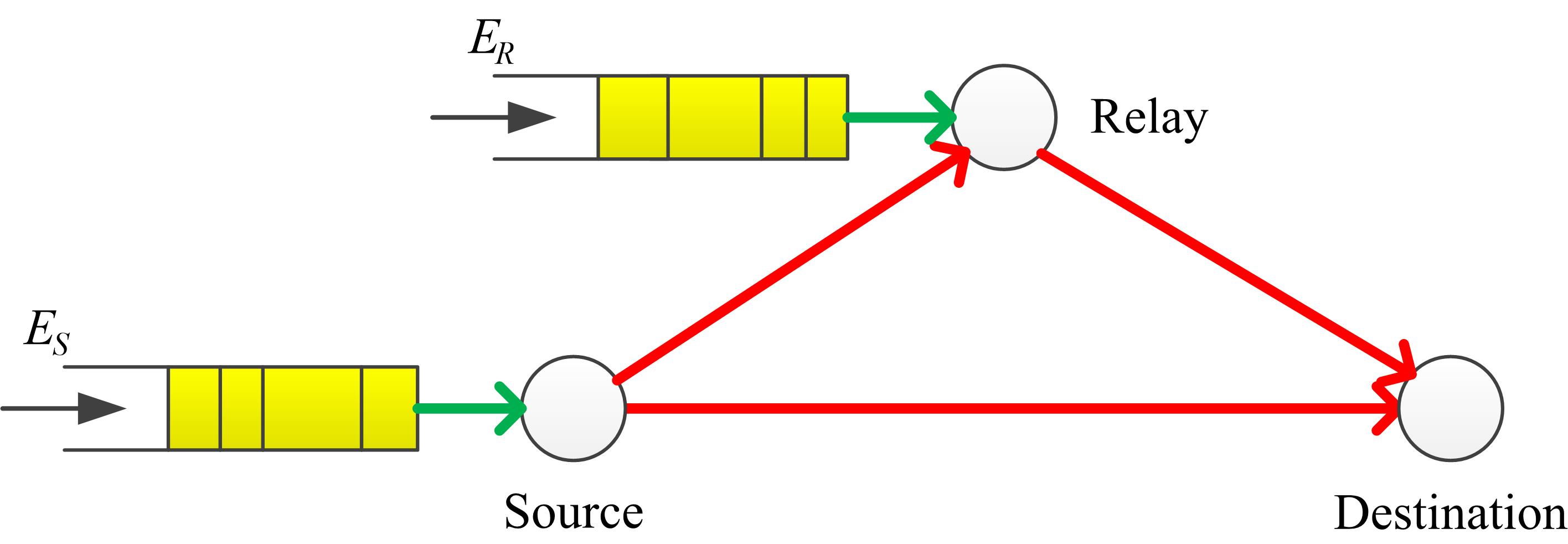}}\\
  \subfigure[Joint power allocation with energy sharing]{\label{withEHsharing}\includegraphics[width=0.50\textwidth]{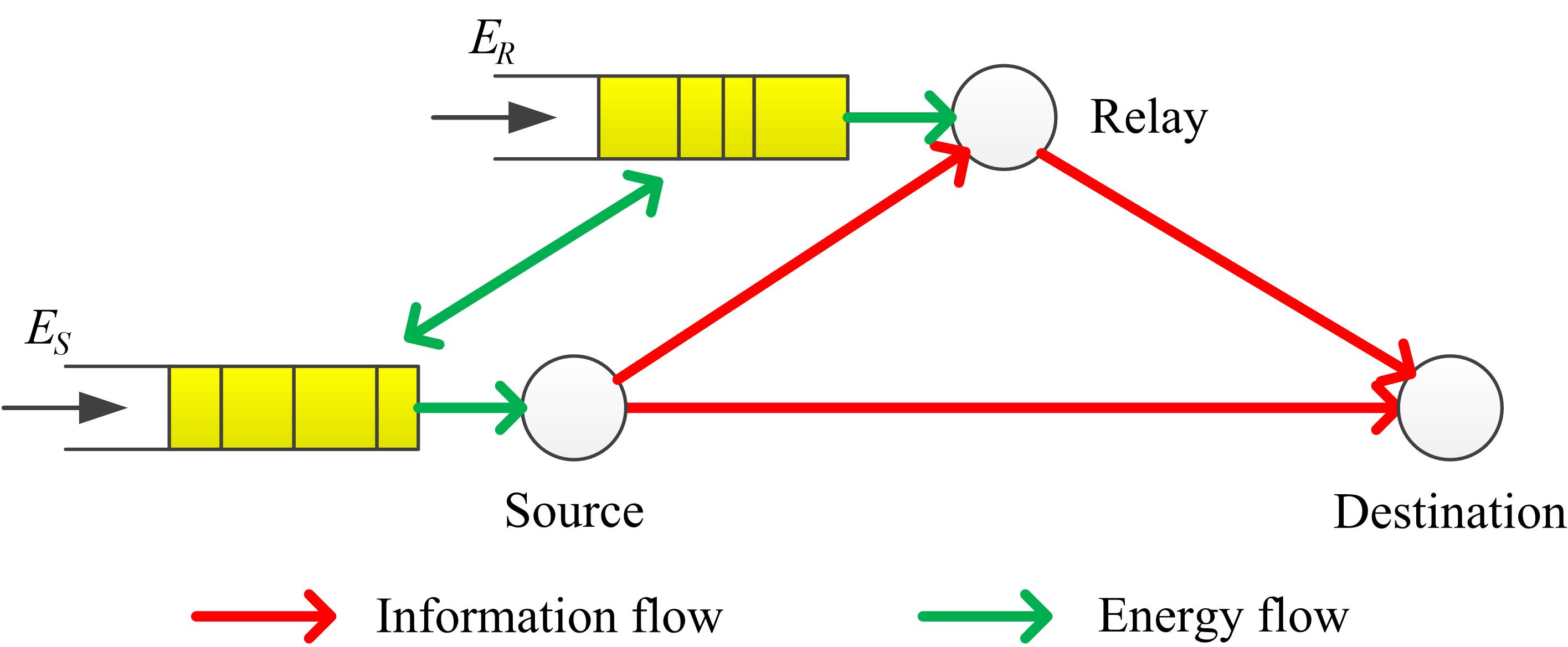}}
  \caption{A three-node relay channel with EH source and relay.}\label{fig:relay}
\end{figure}

We consider a three-node relay channel with half-duplex and orthogonal transmissions, where the relay node transmits (to the destination node) and receives (from the source node) over two different frequency bands. As shown in Fig. \ref{fig:relay}, the source and destination nodes transmit with the power drawn from their own EH devices, while the destination is powered by a constant energy source. Similar to the single-user case, in the $(n,m)$-th communication block, denote $E_{S}(m)$ and $E_{R}(m)$ as the EH rates, $P_{S}(n,m)$ and $P_{R}(n,m)$ as the power allocations at the source and relay nodes, respectively. Then the source and relay nodes are each subject to individual EH constraints similarly as those in the single-user scenario. Note that when the EH rates $\{E_{S}(m)\}$ and $\{E_{R}(m)\}$ are independent (e.g., one uses solar panel and the other uses wind turbine), the source and relay nodes may have very different energy availabilities at each given time. It is thus important to jointly optimize the power allocations at the source and relay nodes based on their available CSI and ESI to maximize the system utility.

For simplicity, in the rest of this subsection, we only consider the end-to-end throughput (from source to destination) of the relay channel shown in Fig. \ref{fig:relay} as the utility function, and furthermore focus on the case with non-causal CSIT and ESIT at both the source and relay nodes (i.e., Case 1 for the single-user setup) with time-invariant (AWGN) channels. We divide our discussions for two scenarios: in the first scenario, the source and relay transmit with their individually harvested energy, while in the second scenario, the source and relay are allowed to share their harvested energy via a new technique so-called wireless energy transfer (WET) (see, e.g., \cite{NiyatoHan2014} and the references therein).

\subsubsection{Joint Power Allocation without Energy Sharing}
This case is shown in Fig. \ref{withoutEHsharing}, where the source and relay nodes adapt their power allocations in a cooperative manner based on their individual EH rates over transmissions to maximize the end-to-end throughput. This problem is solved in \cite{HuangZhangCui2013} by assuming the decode-and-forward operation of the relay. The optimal joint power allocation is shown to depend on whether a new {\it decoding constraint} is imposed by the destination or not, as will be discussed next.

For the delay-constrained traffic, a decoding constraint may be applied at the destination such that the source message needs to be decoded immediately after it is transmitted. In other words, the relay needs to decode and forward its received signal from the source without delay. In this case, the achievable rate for each transmission and relaying is limited by both the available energy at the source and that at the relay at that time. Therefore, the optimal source and relay power allocations are coupled, which could be obtained by a two-dimensional (for both the source and relay) search algorithm \cite{HuangZhangCui2013}.

On the other hand, for the delay-tolerant traffic, the destination could tolerate arbitrary decoding delays provided that all source messages are decoded at the end of the transmission, i.e., the decoding constraint is much more relaxed. Consequently, the relay is allowed to store its decoded source messages and forward them to the destination at a later time based on its energy availability. Due to the relaxed decoding constraint, it is shown in \cite{HuangZhangCui2013} that the optimal power allocations at the source and relay could be decoupled and separately optimized. For both the delay-constrained and delay-tolerant cases, the optimal source and relay power allocations should follow a non-decreasing staircase \cite{HuangZhangCui2013}, which can be considered as an extension of the one-dimensional (for source only) case for the point-to-point EH system discussed in Section III-A.

\subsubsection{Joint Power Allocation with Energy Sharing}

Energy sharing via WET is a new approach that allows different EH nodes to exchange energy between each other with certain losses, such that the nodes with excessive energy can share part of their energy to other nodes with insufficient energy, in order to balance the energy availabilities among the nodes for further improving the system utility.

As shown in Fig. \ref{withEHsharing}, the energy sharing capability offers the source and relay nodes more flexibility to manage their energy usage. For example, when the source node is excessive in energy but the relay node has insufficient energy, the source could either increase its own transmit power to boost the achievable rates of both the source-relay and source-destination links, or transfer a portion of its energy to the relay (with certain energy transfer losses) to increase the rate of the relay-destination link, and vice versa. In general, the joint power allocation with additional energy sharing between the source and relay could improve the throughput. In \cite{Gurakan2013}, by assuming a two-hop relay channel (i.e., ignoring the source-destination link in Fig. \ref{withEHsharing}) with one-way energy sharing from the source to the relay, the KKT optimality conditions are used to find the optimal control of the transmit power levels at the source and the relay, as well as the energy transferred from the source to the relay, to maximize the end-to-end throughput.

\subsection{Other Setups}

In fading channels, the source and relay should use adaptive power allocations to explore the channel variations as well as different energy availabilities among the nodes. In this case, it is anticipated that energy sharing could achieve higher gains than in the time-invariant channel case discussed in the above. On the other hand, when different CSI and/or ESI availability (e.g., causal vs. non-causal) cases are considered at the source and relay, the optimal power allocation (e.g., for throughput maximization or outage probability minimization) becomes far more complicated than the point-to-point case discussed in Section III, with many open questions unsolved yet.

In addition to the simple three-node relay channel, other multiuser systems, such as the multiple-access channel \cite{Gurakan2013}, the two-way channel \cite{Gurakan2013}, and the coordinated multi-point (CoMP) systems \cite{XuGuoZhang2013} with distributed EH based transmitters, have been investigated with or without energy sharing. In the EH based CoMP systems \cite{XuGuoZhang2013}, a new joint communication and energy cooperation approach is proposed, in which distributed EH powered base stations cooperatively design their transmitted signals and the energy amounts exchanged among each other to maximize the downlink sum throughput. Instead of using WET, the authors in \cite{XuGuoZhang2013} proposed to implement energy sharing between base stations by leveraging the smart power grid infrastructure with bidirectional power transfer, which in general has much higher efficiency than WET.

\section{Extensions and Future Research Directions}
Besides the above studies on EH based wireless communications, there have been numerous other valuable works on this new and high-potential subject, which are not discussed in this article due to the page limitation. In this section, we briefly present some recent extensions in this area as well as some promising directions for future work.

\subsubsection{Case with EH Receiver}
Most existing studies on EH based communications have considered EH at the transmitter side only, by assuming that the receivers are powered by a stable energy supply. When the receiver is powered by EH, a similar EH constraint like in the EH transmitter case needs to be applied, but with a key difference that the energy used at the receiver is for decoding the signal instead of sending it at the transmitter. As a preliminary work along this new direction, \cite{Yates2014} showed that the detection and decoding operations dominate the energy cost for EH receivers, and the energy cost is nondecreasing over both the sampling rate and the decoding complexity. Thus, the communication rate should be designed by taking into account the energy availability at both the EH transmitter and EH receiver.

\subsubsection{Cross-Layer Design}
So far, we have been focusing on the physical (PHY) layer design issues on EH based communications by assuming the presence of backlogged data for transmission at all time. However, in practical systems, the data arrives at the transmitter with random timing and amounts in general. In such cases, the transmitter needs to deal with the uncertainties in both energy and data arrivals, and it is thus beneficial to jointly schedule the energy usage and data packet transmission based on the channel conditions \cite{Gunduz2014}. As another example, consider a wireless network with multiple EH transmitters sharing the same limited channel resources for communications, for which there is a necessity for the design of energy-aware medium access control (MAC) to optimize the system throughput. A preliminary work on this problem is presented in \cite{LiHang2014B}, where a distributed opportunistic scheduling scheme is proposed to jointly design the access control and power management for EH transmitters. In general, a cross-layer design approach should be further investigated to achieve more efficient operation of EH based communication systems.

\subsubsection{Hybrid Energy Sources and Imperfect Energy Storage Devices}
Due to the random and intermittent characteristics of practical EH sources, using renewable energy alone may not be sufficient to provide reliable operation of wireless systems with large power demands, e.g., in base stations. To maintain their reliable operations, it is wise to use hybrid energy sources by efficiently integrating the renewable energy with the conventional energy (such as fuel generators). On the other hand, energy storage devices (ESDs) with imperfect charging-discharging efficiency and a finite capacity may be employed in the system. In general, how to optimally design the energy management policies with hybrid energy sources and/or imperfect ESDs to achieve the maximum utility in EH based communications still remains largely open, while some initial results have been obtained \cite{ChiaZhangNew}.

\subsubsection{RF EH with Dedicated WET}
In addition to the conventional EH sources such as wind and solar power as well as ambient radio frequency (RF) transmissions, deploying dedicated power transmission nodes in the network for delivering controllable energy over the air to distributed communication devices (e.g., sensors) has drawn growing interests recently. The devices can either harvest RF energy from the signal transmitted by the power transmission nodes, or decode the information in it, or even use part of the energy harvested to decode the information and the remaining energy to transmit or relay other information \cite{ZhangHo2013}. The RF signal enabled WET is a very promising technique for powering low-power wireless communication devices such as those in sensor networks and personal/body area networks, even with its practically limited energy transfer efficiency, which actually could be alleviated by some new techniques such as highly directional massive MIMO \cite{ZhangHo2013}. Clearly, such WET powered communication brings a new avenue for the future research of EH based systems.

\section{Conclusions}
We have introduced a general utility optimization framework for energy harvesting (EH) based wireless communication systems subject to a new type of energy usage constraint. Under this framework, the solutions for a variety of design problems of high practical interests have been discussed, including throughput maximization and outage probability minimization under different practical assumptions on the channel and energy state information. A new design paradigm for multiuser EH communication systems with joint energy and communication cooperation is also discussed. Promising research directions for extensions are also highlighted. We hope that this article will provide a timely overview of the state-of-the-art results on the fundamental design principles for EH based wireless communications, and serve as an inspiring key that leads to more fruitful results in future works.


\section*{Biographies}
\begin{biographynophoto}
{Hang Li}(S'13) received the B.E. and M.S. degrees from Beihang University, Beijing, China, in 2008 and 2011, respectively. He is currently a Ph.D. student at Texas A\&M University. His current research interests include energy-harvesting-based communications, energy-aware scheduling, and applications of optimal stopping theory and dynamic programming.
\end{biographynophoto}

\begin{biographynophoto}
{Jie Xu} (S'12-M'13) received the B.E. and Ph.D. degrees in electronic engineering and information science from the University of Science and Technology of China, Hefei, China, in 2007 and 2012, respectively. He is currently a Research Fellow with the Department of Electrical and Computer Engineering, National University of Singapore, Singapore. His current research interests include smart grid, multiuser MIMO, energy efficiency, and energy harvesting in wireless communication.
\end{biographynophoto}

\begin{biographynophoto}
{Rui Zhang} (S'00-M'07)  received his Ph.D. degree from Stanford University in 2007. He is now an Assistant Professor with the ECE Department of National University of Singapore. His current research interests include multiuser MIMO, cognitive radio, energy-efficient and energy-harvesting-based wireless communications, and wireless information and power transfer. He was the recipient of the 6th IEEE ComSoc Asia-Pacific Best Young Researcher Award in 2011. He is now an editor for the IEEE Transactions on Wireless Communications and IEEE Transactions on Signal Processing.
\end{biographynophoto}

\begin{biographynophoto}
{Shuguang Cui} (S'99-M'05-SM'12-F'14) received his Ph.D from Stanford in 2005. He is now an associate professor at TAMU. His current research interest is data oriented large-scale information analysis and system design. He was selected as the Thomson Reuters Highly Cited Researcher and listed in the Worlds' Most Influential Scientific Minds by Sciencewatch in 2014. He was the recipient of the IEEE Signal Processing Society 2012 Best Paper Award. He is an IEEE Fellow and ComSoc Distinguished Lecturer.
\end{biographynophoto}

\newpage
\clearpage

\newpage
\clearpage

\newpage
\clearpage

\newpage
\clearpage

\newpage
\clearpage

\end{document}